# Calibration of piezoelectric positioning actuators using a reference voltage-to-displacement transducer based on quartz tuning forks

**Piezoelectric tuning fork as a calibration tool**

*Andres Castellanos-Gomez[1,+,\*], Carlos R. Arroyo[1,+], Nicolás Agraït[1,2,3] and Gabino Rubio-Bollinger[1,2,\*]*.

[1] Departamento de Física de la Materia Condensada (C–III).
Universidad Autónoma de Madrid, Campus de Cantoblanco, 28049 Madrid, Spain.
[2] Instituto Universitario de Ciencia de Materiales "Nicolás Cabrera".
Universidad Autónoma de Madrid, Campus de Cantoblanco, 28049 Madrid, Spain.
[3] Instituto Madrileño de Estudios Avanzados en Nanociencia
IMDEA-Nanociencia, 28049 Madrid, Spain.
[+] Present address: Kavli Institute of Nanoscience, Delft University of Technology, Lorentzweg 1, 2628 CJ Delft, The Netherlands.

E-mail: a.castellanosgomez@tudelft.nl , gabino.rubio@uam.es

*We use a piezoelectric quartz tuning fork to calibrate the displacement of ceramic piezoelectric scanners which are widely employed in scanning probe microscopy. We measure the static piezoelectric response of a quartz tuning fork and find it to be highly linear, non-hysteretic and with negligible creep. These performance characteristics, close to those of an ideal transducer, make quartz transducers superior to ceramic piezoelectric actuators. Furthermore, quartz actuators in the form of a tuning fork have the advantage of yielding static displacements comparable to those of local probe microscope scanners. We use the static displacement of a quartz tuning fork as a reference to calibrate the three axis displacement of a ceramic piezoelectric scanner. Although this calibration technique is a non-traceable method, it can be more versatile than using calibration grids because it enables to characterize the linear and non-linear response of a piezoelectric scanner in a broad range of displacements, spanning from a fraction of a nanometer to hundreds of nanometers. In addition, the creep and the speed dependent piezoelectric response of ceramic scanners can be studied in detail.*

## 1. INTRODUCTION

Since the invention of the scanning tunneling microscope (STM) [1] and the atomic force microscope (AFM) [2], most of the scanning probe techniques make use of piezoelectric transducers to scan a tip over a surface with subatomic resolution. An accurate calibration of the displacements of these transducers as a function of the applied voltage is necessary to obtain quantitative information about the topography of the surface under study which has motivated the development of several procedures to calibrate the *xyz* displacement of the piezoelectric scanners such as optical interferometry [3], imaging calibration specimens [4, 5] and others [6-8].

In this work we present a method to calibrate the displacement of piezoelectric actuators using the static piezoelectric deformation of quartz tuning forks. Although this calibration can be done by simply imaging a calibrated grid, the non-linear behaviour of the piezoscanners makes necessary to use several tens of calibrated gratings to obtain the voltage-to-displacement calibration of the piezoelectric scanner. As the presented method relies on the comparison between the displacement of the uncalibrated actuator and that of a calibrated quartz tuning fork, which is used as a reference, the calibration for a broad range of displacements can be easily obtained. Moreover, unlike imaging calibration grids, this method enables to calibrate the displacements in z-direction for microscopes with limited or even without *xy*-scanning capabilities, like the ones used in some scanning tunneling microscopy break junction experiments [9, 10] or some nanoindenter setups.

We find quartz tuning forks very adequate to be used as reference actuators because they meet most of the desirable properties of an ideal voltage-to-displacement transducer: low aging, hysteresis and creep and high linearity and thermal stability. Moreover the tuning fork geometry guarantees displacements of the reference voltage-to-distance transducer in an appropriate range (0.1 nm – 100 nm) to calibrate the actuators used in scanning probe microscopy (SPM). We have applied the technique presented here to characterize the *xyz* displacements of the piezotube scanner of an AFM.

## 2. MATERIALS AND METHODS

While the primary application of quartz tuning forks is their use as piezoelectric resonators we characterize their static deformation, and obtain their voltage-to-displacement calibration, in order to determine their adequacy as displacement standards to calibrate the ceramic piezoelectric scanner of local probe microscopes.

*Characterization of the static displacement of quartz tuning forks*

Ceramic piezoelectric actuators suffer from severe accuracy limitations for their use as voltage-to-displacement transducers such as the temperature dependence of their piezoelectric properties [11], aging, nonlinear voltage-to-displacement response, hysteresis and creep [6]. These undesirable properties related to their ceramic structure are not expected for single-crystal piezoelectric actuators [12, 13]. Mass production of quartz single-crystal actuators by the electronic and the watch industries makes them very reliable, affordable and easy to obtain. Single crystal quartz, however, has a low piezoelectric constant and thus the shape of the actuator has to be carefully engineered to ensure large enough displacements in order to make the actuator useful to calibrate scanning probe microscopes (SPMs). Quartz tuning forks, present in most of wristwatches, yield displacements in the range from the sub-nanometer to hundreds of nanometers and therefore can be used as reference voltage-to-displacement transducers to calibrate SPMs.

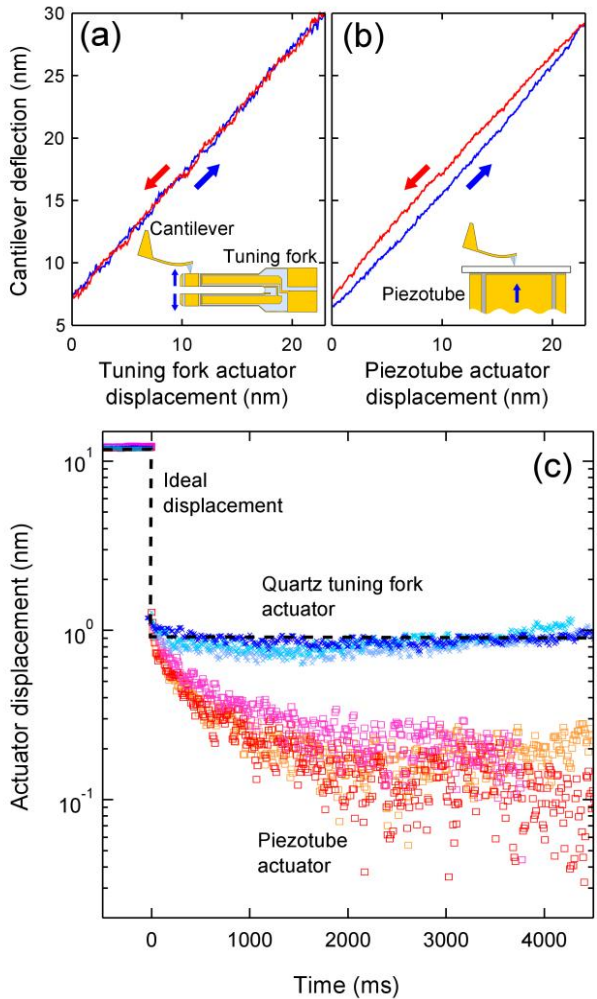

**Figure 1. a:** Cantilever deflection as a function of the *z* axis displacement of the quartz tuning fork transducer prong. The arrows indicate the direction of motion **b:** Cantilever deflection *vs*. the *z* axis displacement of the microscope piezoelectric tube scanner. **c:** Actuator displacement *vs*. time traces after the excitation voltage is suddenly changed. While the quartz tuning fork the displacement closely follows the ideal one (dashed line), the piezotube actuator continues its displacement due to creep.

While quartz tuning forks are widely used in local probe microscopy as high quality factor resonators to measure atomic scale forces [14-18] they can also be employed as static piezoelectric displacement transducers when a DC voltage is applied between the tuning fork electrodes. To characterize the capabilities of quartz tuning forks as piezoelectric transducers we use a position sensor based on an AFM setup where the tuning fork is mounted in the sample holder. Once the tip of a cantilever is in contact with the end of the tuning fork prong, the cantilever deflection follows the motion of the tuning fork. The cantilever deflection is measured fast and accurately by means of a laser beam deflection setup. Fig. 1 shows the deflection of an AFM cantilever in contact with the tuning fork prong as it is displaced in the *z* direction (a) or alternatively using the piezoelectric tube actuator of the microscope (b). The cantilever deflection *vs*. displacement trace measured using the tuning fork actuator shows high linearity and negligible hysteresis (Fig. 1a). In contrast, the motion of the microscope scanner displays a noticeable non-linearity and creep (Fig. 1b). Another benchmark to characterize the motion of piezoelectric actuators is to follow the time evolution of the displacement after a sudden change of the applied voltage (Fig. 1c). We find that the displacement of quartz tuning fork actuators closely follows the ideal step-like displacement whereas the microscope scanner transducer suffers from a strong creep, showing a continued displacement over time despite the applied voltage is fixed. This adverse behavior, typical of piezoelectric ceramics at room temperature, is undesirable for experiments in which it is necessary to perform accurate displacements followed by measurement intervals without any movement such as spectroscopic measurements carried out at different tip-sample distances.

*Calibration of the displacement of a quartz tuning fork*

In order to use the quartz tuning fork as reference voltage-to-displacement transducer one has to determine with accuracy the deflection of the tuning fork prongs ($\Delta z$) produced by a DC voltage ($V_{DC}$) applied to its electrodes. In previous works this calibration of the tuning fork was done with an optical interferometer [13]. This technique, however, requires a dedicated setup capable of measuring deflections in the nanometer range. An alternative approach can be used taking advantage of the high quality factor $Q$ of the mechanical resonance of the tuning fork whose amplitude ($A$) can be as large as several microns for an excitation voltage ($V_{exc}$). As the tuning fork behaves effectively as a harmonic oscillator [19] one can relate the motion of the prongs at resonance to their static deflection as:

$$\frac{\Delta z}{V_{DC}} = \frac{1}{Q}\frac{A}{V_{exc}}. \tag{1}$$

Note that although the $Q$ factor can vary with the ambient conditions such as the humidity, the product $(A \cdot V_{exc}^{-1})Q^{-1}$ and thus the DC calibration of the tuning fork does not change. This is because the DC calibration of the tuning fork only depends on the piezoelectric properties of the quartz and the geometry of the tuning fork and not on the ambient conditions. To ensure a right tuning fork calibration, nevertheless, both the Q factor and the ratio $A \cdot V_{exc}^{-1}$ should be measured under the same ambient conditions. This approach is convenient because of the large variety of methods developed to measure the oscillation amplitude of the tuning fork prongs [20-22]. Here we opt for the optical inspection of the oscillation using a Canon EOS 550D camera attached to a Nikon Eclipse LV-100 optical microscope. We have found that the accuracy of the measurement of the oscillation amplitude improves using stroboscopic illumination [19]. To do so, the tuning fork is illuminated with a light emitting diode modulated at twice the excitation frequency. We adjust the phase shift between the illumination and the excitation voltage to obtain an image that is the superposition of the two extremals of an oscillation cycle. Fig. 2b-d shows an example of three optical micrographs of the end of a tuning fork prong under stroboscopic illumination (Fig. 2a) acquired for increasing excitation voltage amplitude. The relationship between the oscillation amplitude of the prongs and the excitation voltage (Fig. 2f) is linear with a slope of $5496 \pm 60$ nm/V. The inset in Fig. 2f shows the measured resonance spectrum from which one can obtain a quality factor $Q = 4667 \pm 3$. Using expression (1) the calibration of the tuning fork used as reference transducer is $\Delta z/V_{dc} = 1.177 \pm 0.025$ nm/V. Although this calibration method is not traceable, its validity can be checked. This has been done by comparing the DC calibration determined with the method proposed here with the one that we have experimentally measured with an optical interferometer [13]. It is also important to emphasize that once a tuning fork is calibrated it can be used for several years because of the low aging of the quartz piezoelectric properties.

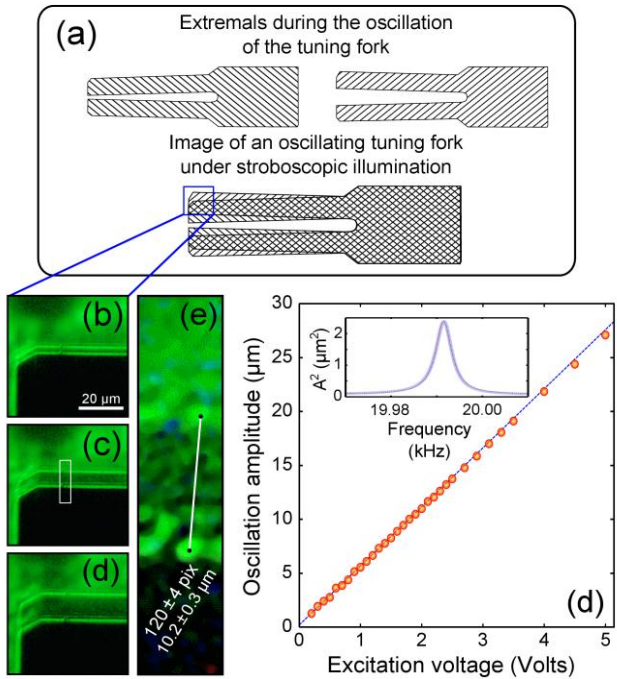

**Figure 2. a:** Schematic diagram of an image of an oscillating quartz tuning fork under stroboscopic illumination. The image is the superposition of two oscillation instants phase shifted by 180º. **b-c:** Optical micrographs of the end of a tuning fork prong under stroboscopic illumination with excitation voltage of $V_{exc}$ = 0.7, 1.8 and 3.2 V . **e:** Detail of the region marked with a rectangle in **c**. **d:** Measured oscillation amplitude of the prong as a function of the excitation voltage and a linear fit (dashed line). **Inset in e:** Resonance spectrum and a Lorentzian fit (dotted line).

## 3. RESULTS

We use a calibrated tuning fork as a reference displacement transducer to characterize the three axis (*xyz*) displacement of the piezotube scanner of a local probe microscope. The tuning fork is mounted as a sample and its calibrated displacements are monitored by means of the feedback control loop of the microscope. In particular, we perform the calibration of the piezotube scanner of a Nanotec Electronica Cervantes AFM system operated in amplitude modulation AFM mode (AM-AFM mode). The calibration in the *z* axis is obtained by measuring the response of the feedback control loop for a well-defined displacement of the tuning fork in the *z* direction. The calibration in the *x* or *y* direction is determined from the analysis of the shift of a topography profile when the tuning fork is deflected in the *x* or *y* direction.

*Calibration of the z-displacement of a piezotube scanner*

To calibrate the *z* displacement the reference tuning fork is mounted on the AFM sample holder so that the displacement of the prongs is in the *z* direction (Fig. 3a). The AFM tip can be readily positioned at the end of the tuning fork prong with the help of a long working distance optical microscope attached to the AFM and two micrometer screws that change the *xy* position of the cantilever with respect to the sample. Note that most commercial AFMs have *xy* positioning capability as well as an optical microscope. Then the amplitude modulation AFM mode (AM-AFM mode) is used to keep the tip-prong distance constant. To determine the $z$ axis calibration we record the voltage, applied to the $z$ electrode of the piezotube, necessary to keep constant the tip-prong distance while we establish a well-defined deflection of the tuning fork prong. The relationship between the prong deflection and the piezotube voltage is rather linear for a range of ± 5 nm. From the slope of this relationship we determine the $z$ axis calibration of the piezotube $10.6 \pm 0.6$ nm/V. This value is in good agreement with the calibration $11 \pm 1$ nm/V obtained by measuring the height of monoatomic steps in graphite (HOPG).

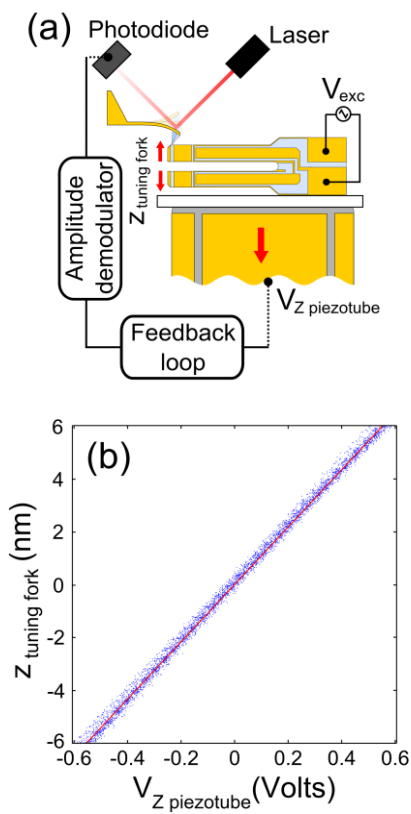

**Figure 3. a:** Schematic diagram of the experimental setup used to calibrate the $z$ axis displacement of the piezotube scanner of an AFM. **b:** Relationship between the deflection of the tuning fork prong and the voltage applied to the $z$ sector of the piezotube to keep the tip-prong distance constant.

A more detailed analysis of the residues of the linear fit shown in Fig. 3.b reveals that even for these moderate displacements of the piezo there is a non-negligible non-linear behaviour. A quadratic fit to the data gives a calibration for the piezo $\Delta z(V) = -0.31V^2 + 10.6V$ (with $\Delta z$ in nm and $V$ in volts).

In the configuration shown in Fig. 3.a., the tuning fork can be also used to adjust the feedback parameters of the microscope. This can be done by monitoring the output of the feedback loop while a low frequency (1-10 Hz) triangular-shape signal is applied to the tuning fork electrodes. Then, both proportional and integral gains can be adjusted until the feedback output reliably follows the triangular shape of the tuning fork displacement. By applying a square signal instead of a triangular signal the time-response and the creep of the piezo scanner can be characterized.

*Calibration of the xy-displacement of a piezotube scanner*

The calibration of the *xy* displacement of the piezotube scanner is carried out mounting the reference tuning fork on the AFM sample holder with the axis of the deflection of the prongs parallel to the *x* displacement of the piezotube (Fig. 4a). The AFM tip is positioned at the end of one prong of the tuning fork and the AM-AFM mode is used to keep constant the tip-prong distance. The *x* axis calibration of the piezotube can be obtained by acquiring a topographic profile of the prong surface before and after establishing a well defined displacement of the tuning fork prong. Fig. 4b shows two consecutive topographic profiles measured before and after a displacement of the prong of $\Delta d = 23.55$ nm. The misalignment between the movement of the piezotube scanner and the prong of the tuning fork has been also taken into account, resulting in an *x* displacement of the prong $\Delta x = \Delta d \cdot \sin(85º) = 23.46$ nm. The two profiles are shifted by $0.85 \pm 0.02$ V in the voltage applied to the *x* electrode of the piezotube which gives a calibration of the piezotube of $27.6 \pm 0.7$ nm/V. We find that this calibration is in agreement with the one obtained by imaging a microfabricated AFM calibration grating[1] ($28 \pm 1$ nm/V).

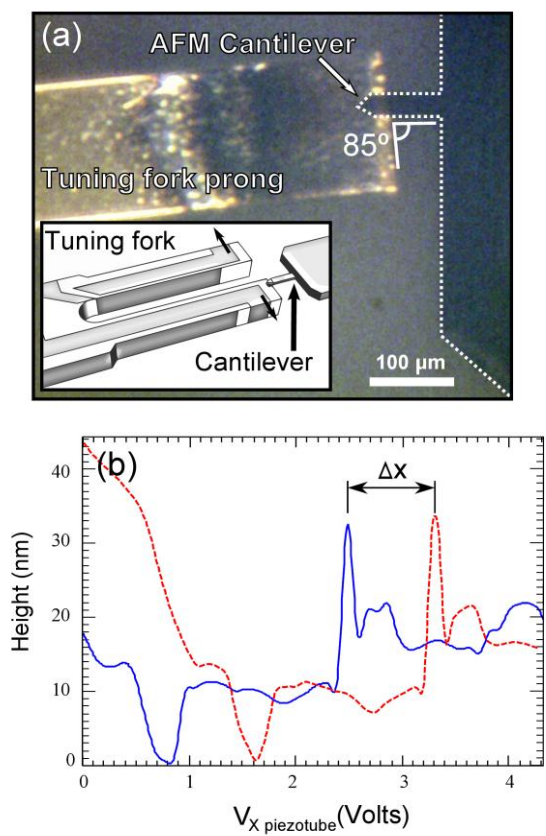

**Figure 4. a:** Optical micrograph of the end of a tuning fork prong with the AFM cantilever positioned on top. The small misalignment between the axis of the deflection of the prongs and the *x* axis of the piezotube has been taken into account to calibrate the piezotube. **Inset in a:** Schematic of the tuning fork arrangement used to calibrate the *x* displacement of the AFM piezotube. **b:** Two consecutive topographic profiles of the prong surface measured before (solid blue trace) and after (dashed red trace) establishing a static deflection $\Delta x = 23.46$ nm of the tuning fork prong.

---

[1] Purchased from nanoScience Instruments (part number 32400). It is a chessboard-like structure with 2 μm square pattern with a 4 μm period.

In order to study the non-linear response in *xy* of the piezoelectric scanner, one can repeat the process described above with increasing displacements of the tuning fork prongs. The non-linear response of the piezo scanner can be obtained from the relationship between the displacements of the tuning fork prong and the apparent shifts of the topographic profiles. Another possibility would be recording an AFM image without movement in the slow scan axis while a low-frequency triangular wave is applied to the tuning fork electrodes. In this way one would have a collection of topographic line profiles shifted from each other following a triangular shape form. From these profiles one can extract the voltage-to-displacement relationship in the *x* or *y* directions for a broad range of displacements.

## 4. CONCLUSIONS

We have characterized the static displacement of piezoelectric quartz tuning forks in order to determine their adequacy as reference voltage-to-displacement transducers. We have found that, due to their single-crystalline structure, quartz tuning forks perform like ideal piezoelectric transducers. Their high voltage-to-displacement linearity and low hysteresis and creep make these quartz actuators superior to conventionally used piezoelectric ceramic actuators. Additionally, quartz actuators in the form of tuning fork can yield displacements in the range required in scanning probe microscopes (0.1 nm to 100 nm). We have presented a method to calibrate piezoelectric scanners, used in scanning probe microscopy, by comparing the displacement of an uncalibrated actuator with one of a calibrated tuning fork. We have applied this versatile method to characterize the three axis displacement of the piezotube scanner of a commercial AFM. There are advantages employing a tuning fork as a calibrated voltage-to-displacement reference instead of using calibration grids. On the one hand, the displacement range can be selected, according to the size of the structures to be characterized, in a broad range spanning from a fraction of a nanometer to hundreds of nanometers. On the other hand, a detailed study of the ceramic piezoelectric scanner creep and the speed dependent calibration can be performed for experiments that require a precise positioning of the tip of a microscope. In addition, the tuning fork itself can be used as a supplement in a scanning probe microscope to perform well controlled displacements without noticeable creep.

## 5. ACKNOWLEDGEMENTS

A.C-G. acknowledges fellowship support from the Comunidad de Madrid (Spain). This work was supported by MICINN (Spain), MAT2008-01735 and CONSOLIDER-INGENIO-2010 'Nanociencia Molecular' CSD-2007-00010 and Comunidad de Madrid "Nanobiomagnet" S2009/MAT-1726.